\documentclass{www2010-submission}
\usepackage{url}
\begin{document}

\title{An HTTP-Based Versioning Mechanism for Linked Data}

\numberofauthors{3}
%


\author{
%
\alignauthor Herbert Van de Sompel\\
       \affaddr{Los Alamos National Laboratory,}
       \affaddr{NM, USA}\\
       \email{herbertv@lanl.gov}
\alignauthor Robert Sanderson\\
       \affaddr{Los Alamos National Laboratory,}
       \affaddr{NM, USA}\\
       \email{rsanderson@lanl.gov}
\alignauthor Michael L. Nelson \\
       \affaddr{Old Dominion University,}
       \affaddr{Norfolk, VA, USA}\\
       \email{mln@cs.odu.edu}
\and
\alignauthor Lyudmila L. Balakireva\\
       \affaddr{Los Alamos National Laboratory,}
       \affaddr{NM, USA}\\
       \email{ludab@lanl.gov}
\alignauthor Harihar Shankar \\
       \affaddr{Los Alamos National Laboratory,}
       \affaddr{NM, USA}\\
       \email{harihar@lanl.gov}
\alignauthor Scott Ainsworth \\
       \affaddr{Old Dominion University,}
       \affaddr{Norfolk, VA, USA}\\
       \email{sainswor@cs.odu.edu}
}

\maketitle
\begin{abstract}

Dereferencing a URI returns a representation of the current state of the resource identified by that URI. But, on the Web representations of prior states of a resource are also available, for example, as resource versions in Content Management Systems or archival resources in Web Archives such as the Internet Archive. This paper introduces a resource versioning mechanism that is fully based on HTTP and uses datetime as a global version indicator. The approach allows ``follow your nose'' style navigation both from the current \textit{time-generic} resource to associated \textit{time-specific} version resources as well as among version resources. The proposed versioning mechanism is congruent with
the Architecture of the World Wide Web, and is based on the Memento framework that extends HTTP with transparent content negotiation in the datetime dimension. The paper shows how the versioning approach applies to Linked Data, and by means of a demonstrator built for DBpedia, it also illustrates how it can be used to conduct a time-series analysis across versions of Linked Data descriptions.

\end{abstract}


\section{Introduction}
\label{sec:intro}
The Architecture of the World Wide Web \cite{archWWW} states that
dereferencing a URI yields a representation of the (current) state of the
resource identified by that URI, and highlights the impracticality of keeping
prior states accessible at their own distinct URIs:

\begin{quote}
Resource state may evolve over time. Requiring a URI owner to publish a new
URI for each change in resource state would lead to a significant number of
broken references. For robustness, Web architecture promotes independence
between an identifier and the state of the identified resource.
\end{quote}

Nevertheless, use cases abound that require the availability of
(representations of) distinct prior states of resources. Resource
versioning is of crucial importance in areas as diverse as
community-driven content creation, open government, and scientific
communication. Also, as more data becomes available in the Linked Data
cloud, the need to version them will increase if only to allow efficient
update of stores that leverage the data, and to trace their provenance.
Web archives and content management systems possess significant amounts
of prior versions of resources, but these prior versions are largely
disconnected from current versions and discoverable only in an ad-hoc
manner.  Given this state of Web resource versioning, we consider these
challenges:


\begin{enumerate}

\item \textit{Given the current version of a resource, how can ``follow
your nose'' style navigation to prior versions of the resource be
achieved?}

\item \textit{Given any version of a resource and a particular timestamp,
how can ``follow your nose'' style navigation towards another version
that matches the timestamp be achieved?}

\end{enumerate}

This paper is concerned with versioning mechanisms that are
\textbf{machine-actionable}, have \textbf{global scope}, and are
\textbf{independent of media-type}. Hence, approaches that are mainly
beneficial to human users such as untyped hyperlinks in HTML with
anchor text that provides navigational guidance (e.g. ``previous/next
version''), or version semantics expressed in metadata-carrying URIs
\cite{tag:MetadataInURIs} are not considered. Also, mechanisms that have
version indicators specific to a certain server such as the deprecated
``Content-Version'' header field from RFC 2068 \cite{rfc2068} are
not considered.  Similarly, versioning mechanisms that are specific
to media-types such as the \textit{link} element combined with the
``prev'' and ``next'' relationships as used in HTML \cite{html:spec}
or Atom \cite{rfc:4287} are not considered.  


Our contributions are a resource versioning 
mechanism based on the global notion of time and an HTTP-based
mechanism to navigate across versions.  Furthermore, we demonstrate
how these contributions can be applied for time series analysis 
across resource versions, and illustrate this using a linked data example.

The remainder of this paper is structured as follows: Section
\ref{sec:classicversion} discusses and illustrates characteristics of resource
versioning approaches; Section \ref{sec:memento} provides an introduction
to the Memento framework that extends HTTP with transparent datetime content
negotiation capabilities; Section \ref{sec:mementoversion} shows how
the Memento framework suggests an elegant resource versioning approach
that is fully based on HTTP; Section \ref{sec:linkeddata} shows how the
Memento versioning ideas apply to Linked Data; Section \ref{sec:dbpedia}
describes the demonstrator we built for the DBpedia environment to
illustrate how the proposed versioning mechanism can be used to access
prior descriptions of DBpedia concepts via their existing DBpedia
URIs. In the same section, we also show how this mechanism was used to
conduct a time-series analysis of Gross Domestic Product values for several countries across
DBpedia versions. Section \ref{sec:related} reviews some related work,
and Section \ref{sec:conclusion} holds our conclusions.

\section{Resource Versioning}
\label{sec:classicversion}

This Section introduces core characteristics of versioning approaches,
discusses these characteristics for a typical resource versioning  
approach,
and evaluates how that versioning approach can meet the challenges (1)  
and (2) from the Introduction.

\subsection{Versioning Characteristics}
\label{sec:versioning-characteristics}

The following four core characteristics of versioning approaches are  
considered:

\begin{enumerate}

\item \textit{Identification}: By which means are different versions  
identified?

\item \textit{Versioning Strategy}: What is the approach used to assign  
identifiers to versions,
e.g. do new versions receive a new identifier, do they inherit a prior  
identifier, etc.?

\item \textit{Version Relationships}: How are version relationships between  
resources expressed?

\item \textit{Version Timestamping}: How is the datetime associated with  
versions conveyed?

\end{enumerate}

\subsection{A Typical Resource Versioning Approach}
\label{sec:versioning-typical}

\begin{figure}
\begin{center}
\psfig{file=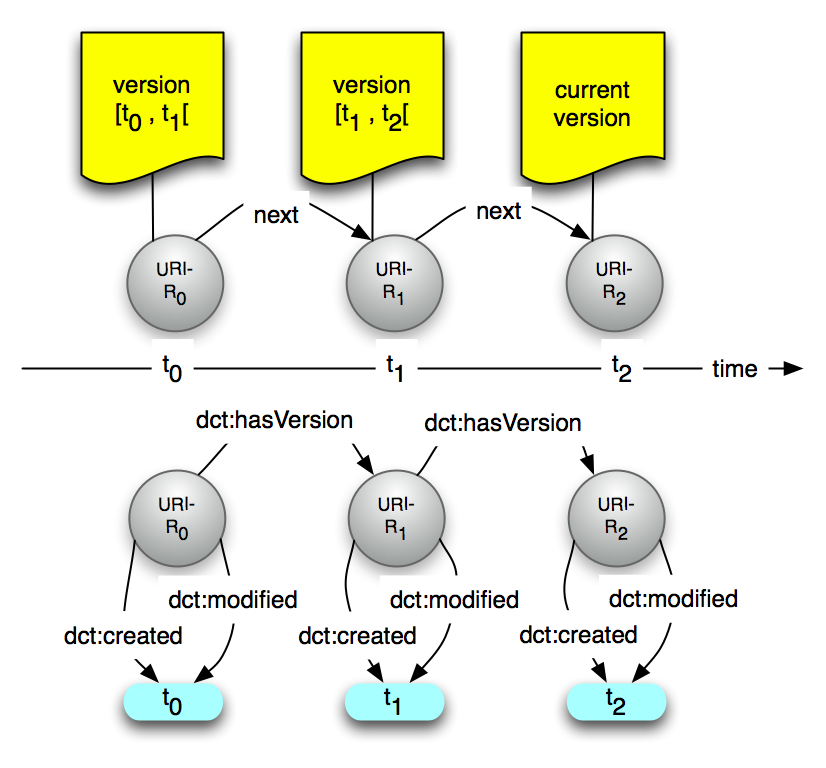, width=3.0in}
\caption{A resource versioning strategy, and its expression in RDF.}
\label{fig:classicversion}
\end{center}
\end{figure}

The quote from the Architecture of the World Wide Web implicitly  
suggests a versioning approach as depicted at the top of Figure 
\ref{fig:classicversion}.
This approach is described in terms of the aforementioned core  
characteristics:

\begin{enumerate}

\item \textit{Identification}:
HTTP URIs are used to identify versions of resources; each version has  
its own URI.

\item \textit{Versioning Strategy}:
A new URI is minted for each for each new version. When a use case
requires that a resource URI-R$_0$ that started its existence at
t$_0$, but at time t$_1$ changes state in such a way that a distinct
identity is needed, a new resource with URI-R$_1$ is minted. And,
if consecutively at time t$_2$ a change in state of URI-R$_1$ again
requires a new identity, a resource URI-R$_2$ is created (top of Figure
\ref{fig:classicversion}). URI-R$_0$, URI-R$_1$, and URI-R$_2$ co-exist
and, in terms of \cite{tbl:generic}
and its associated ontology\footnote{Ontology for Relating Generic and  
Specific Information Resources \url{http://www.w3.org/2006/gen/ont}}  
are \textit{time-specific} resources. They represent the evolving  
state of
a not explicitly defined, abstract resource and are interlinked by the
\url{http://www.w3.org/2006/gen/ont#sameWorkAs} property.

\item \textit{Version Relationships}:
Can be made available as RDF metadata published about and linked from  
the related resources. The common Dublin Core Terms\footnote{\url{http://dublincore.org/documents/dcmi-terms/ 
}} \textit{hasVersion} and \textit{isVersionOf} predicates (bottom of  
Figure \ref{fig:classicversion}) can be used. Alternatively, the  
machine-processable and media-independent \textit{HTTP Link} header  
\cite{rfc:http-link} can be used in combination with the registered  
\textit{prev} and \textit{next} relationships to express version-relationship semantics. 
In both cases, it should be noted that the  
semantics of the relationships do not strictly only apply to time-based version relationships: \textit{hasVersion} is used in cases  
where ``a related resource is a version, edition, or adaptation of the  
described resource'', whereas \textit{next} refers to ``the next  
resource in an ordered series of resources''.

\item \textit{Version Timestamping}:
In case of the aforementioned RDF approach, additional triples can be  
introduced to express version timestamps. For example, at the bottom  
of Figure \ref{fig:classicversion}, the Dublin Core Terms predicates  
\textit{created} and \textit{modified} are used in accordance to the  
description for time-specific resource given in the aforementioned  
ontology as being a resource for which ``the dates of creation and of  
last modification are the same''. It is unclear how a version  
timestamp can appropriately be expressed when using the HTTP Link  
element: conveying such information is not specified for the  
\textit{prev} and \textit{next} relationships, the HTTP Last-Modified  
header does not provide reliable version semantics, and use of  
metadata embedded in the linked resource yields an approach that is  
dependent on media-type.

\end{enumerate}

The above characterization reveals the technological substrates that  
are used in the considered versioning approaches: for the RDF  
approach, URIs, HTTP, RDF, and an appropriate RDF vocabulary; for the  
HTTP Link approach, URIs, HTTP, HTTP Link, and registered link  
relationships. For both approaches, applications such as browser plug- 
ins, can be created to support the navigation described in question  
(1) above, whereby the starting point would be the current resource  
URI-R$_2$. Also (2) can be achieved for the RDF approach, although a  
processor would need to traverse versions until a matching datetime is  
found. Lacking appropriate version datetime information, (2) can not  
reliably be achieved in case of the HTTP Link header.

To summarize, both (1) and (2) can be achieved for the RDF approach,  
however: (a) Two technological substrates, HTTP and RDF, must be  
combined (b) Version datetime can not be used as a primary entry  
point; rather resource versions must be traversed until a version with a  
matching datetime is found (3) The common predicates used to express  
version relationships do not necessarily imply time-based version  
relations.

\section{The Memento Framework}
\label{sec:memento}

The basic motivation for the Memento\footnote{\url{http://mementoweb.org/}} work \cite{nelson:memento:tr}
is achieving a tighter integration between the current and the past
Web. Remnants of the past Web exist both in version-aware servers
such as Content Management Systems (CMS, e.g. Wikipedia) and Version
Control Systems, and in special-purpose Web Archives such as the
Internet Archive\footnote{\url{http://archive.org/}} and the on-demand
WebCite\footnote{\url{http://webcitation.org/}} archive. Whereas a
current representation of a resource is available from its URI-R,
prior representations - if they exist - are available from distinct
resources URI-M$_i$ (i=1..n) that encapsulate the state URI-R had
at times t$_i$, with t$_i$ prior to the current time. In the Memento
framework, the resource that provides the current representation is
named the \textit{Original Resource}, whereas resources that provide
prior representations are named \textit{Mementos}. More formally, a
Memento for a resource URI-R (as it existed) at time t$_i$ is a resource
URI-M$_i$[URI-R@t$_i$] for which a representation at any moment past
its creation time t$_c$ is the same as a representation that was
available from URI-R at time t$_i$, with t$_c$ $\geq$ t$_i$. Implicit
in this definition is the notion that, once created, a Memento always
keeps the same representation.

From a HTTP perspective, URI-R and URI-M$_i$ are disconnected in that
HTTP provides no means to navigate towards a URI-M$_i$ via its original
URI-R. The Memento framework introduces this missing capability as follows
(Figure \ref{fig:memento}):

\begin{figure} 
\begin{center}
\psfig{file=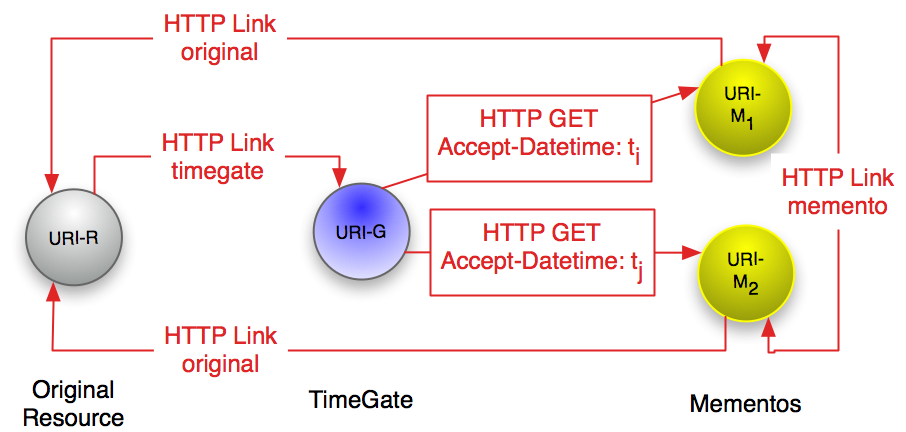, width=3.5in}
\caption{The Memento Framework.} 
\label{fig:memento}
\end{center}
\end{figure}

\begin{itemize}

\item Inspired by Transparent Content Negotiation for HTTP
(\textit{conneg} from now on) specified in RFC 2295 \cite{rfc2295} that
allows HTTP clients to negotiate with HTTP servers in four dimensions
(media type, language, character set, compression), Memento introduces
conneg in a fifth dimension: datetime. RFC 2295 introduces the notion
of a transparently negotiable resource as the resource that is the
target of conneg, and variant resources that vary according to the
aforementioned negotiable dimensions. Similarly, Memento introduces
the notion of a TimeGate URI-G as a resource that supports conneg in
the datetime dimension, and Mementos URI-M$_i$[URI-R@t$_i$] as the
resources that vary according to the datetime dimension. In a manner
symmetrical to the way RFC 2295 introduces the \textit{Accept-Language}
request header to express the client's language preferences, and
the \textit{Content-Language} response header to express the language
returned by the server, Memento introduces the \textit{Accept-Datetime}
and \textit{Content-Datetime} headers to express the client's preferred
datetime for a Memento, and the datetime of the Memento returned by its
hosting server, respectively. It can be noted that, although RFC 2295
did not specify datetime conneg, its desirability is at least suggested
by Tim Berners-Lee's Generic Resources Statement \cite{tbl:generic} as
all other dimensions of genericity described in it (language, media-type,
target-medium) are covered by RFC 2295.

\item In order to support discovery of a TimeGate URI-G for a resource
URI-R, a relationship type of \textit{timegate} is introduced for the HTTP
Link response header \cite{rfc:http-link}. In case of servers that have
internal versioning/archiving support (such as CMS) a TimeGate URI-G for
URI-R can typically be exposed by the server of URI-R itself. In cases
whereby servers rely on third parties for their versioning/archiving
(for example by being recurrently crawled by the Internet Archive),
URI-R and URI-G will reside on different servers. In addition, in order
to allow discovering the Original Resource associated with a Memento,
another special-purpose HTTP Link header, this time with a relationship
type of \textit{original} is introduced.

\item Memento also introduces the notion of a TimeBundle resource via
which an overview is available of all Mementos that a server hosts for a
given (internal or external) URI-R. A TimeBundle is a non-information
resource \cite{coolURIssemanticweb} modeled as an ORE Aggregation
\cite{ldow2009:ore} in which all Aggregated Resources share a temporal
relationship with the Original Resource. A TimeBundle is described by
a TimeMap, which is a specialization of an ORE Resource Map. A TimeMap
lists all URI-M$_i$ for a given URI-R as well as their associated
metadata including timestamp. It also lists the Original Resource URI-R
and its TimeGate URI-G.  Appendix A shows an example RDF/XML TimeMap;
other serializations such as Turtle and Atom are possible.  Discovery of
TimeBundles is supported by the rel value \textit{timebundle} in the
HTTP Link response header.

\end{itemize}

Three aspects of the Memento architecture ensure that the globally
deployed HTTP caching infrastructure can be leveraged.  First, the
Original Resource URI-R and its TimeGate URI-G are always separate
resources: URI-R is a conventional resource and URI-G is dedicated to
datetime conneg.  This eliminates caching problems that would be caused
by transitioning URI-R between non-negotiable and negotiable if URI-R
and URI-G were to coincide.  Second, the initial Memento architecture
\cite{nelson:memento:tr}, required the Original Resource URI-R to
302 redirect to its TimeGate URI-G; as a result, cached versions of
URI-R could not be leveraged.  By using the \textit{timegate} HTTP
Link for discovery of URI-G, Memento clients work with caches instead
of against them.  Third, URI-G and URI-M are never the same resource
so the Mementos (URI-M) can be cached as well.

A detailed overview of HTTP request/response
scenarios is available in the Memento HTTP Transactions
Guide\footnote{\url{http://www.mementoweb.org/guide/http/}}. Here, we
highlight certain aspects related to HTTP interactions with the TimeGate
URI-G.  A choice was made to handle cases in which URI-G is dereferenced
without the Accept-Datetime header, by issuing a ``302 Found'' redirect
to the most recent Memento, as opposed to offering a list of choices to
the client.  While a list would be feasible for a top-level resource (say,
an HTML page), it would be cumbersome for the potentially many embedded
resources (say, the images in the HTML page).  URI-G will only return
HTTP response code ``300 Multiple Choices'' if explicitly requested with a
``Negotiate: 1.0'' request header or when there are multiple Mementos with
the same Content-Datetime\footnote{This may occur as HTTP only supports
second-level time granularity}.  URI-G will return HTTP response code
``406 Not Acceptable'' when the Accept-Datetime is outside of the datetime
range of known Mementos.  For further technical details about the Memento
framework, we refer to the original paper \cite{nelson:memento:tr}, and
the more recent overview of the evolved solution \cite{memento:slides:feb}
that has resulted from feedback to the original ideas provided by both
the Linked Data and Web Archiving communities.

Since its publication, Memento has received significant
attention. Major Web Archives have started implementing
support\footnote{See Agenda of First Memento Implementation Meeting
at \url{http://mementoweb.org/events/IA201002/}}, and work is ongoing
to develop support for common CMS platforms such as MediaWiki and
Drupal. Also, the establishment of a Memento-track at the JISC Developer
Days (Dev8D)\footnote{\url{http://wiki.2010.dev8d.org/w/Talk_6}} organized
by the UK's Joint Information Systems Committee is an early indication
of interest by both funders and implementers.

\begin{figure*}[ht]
\begin{center}
\includegraphics[scale=0.26]{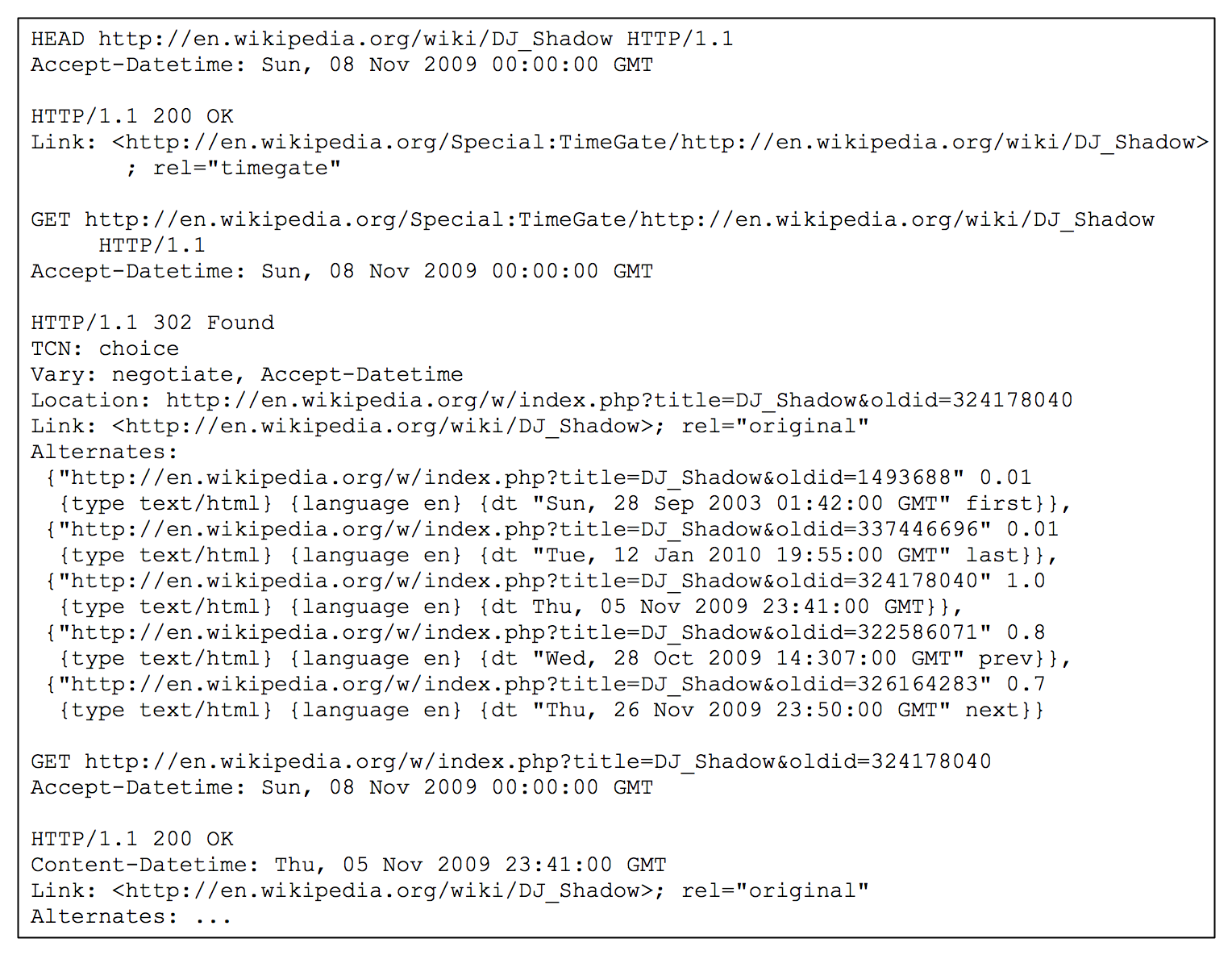}
\caption{Memento HTTP Request/Response Cycle.}
\label{fig:mementoHTTP}
\end{center}
\end{figure*}

As an illustration, Figure \ref{fig:mementoHTTP} shows a Memento HTTP flow whereby a client requests a November 8 2009 version of the Wikipedia page for DJ Shadow, by interacting with its current URI \url{http://en.wikipedia.org/wiki/DJ_Shadow}; the client is pointed by \url{http://en.wikipedia.org/wiki/DJ_Shadow} to a TimeGate at Wikipedia; via that TimeGate the client successfully retrieves a Memento that meets its datetime preferences (only headers crucial to convey an understanding of datetime conneg are shown). We should point out that Wikipedia has not (yet) implemented such Memento HTTP flows, but a MediaWiki plug-in that adds Memento support is available\footnote{Memento MediaWiki plug-in \url{http://www.mediawiki.org/wiki/Extension:Memento}}.

In Figure \ref{fig:mementoHTTP}, note the use of the HTTP Link header to express
the very first and most recent Mementos available from Wikipedia (\textit{rel=``first-memento''} and \textit{rel=``last-memento''}, respectively) as well as the Mementos that are closest in time (\textit{rel=``prev-memento''} and \textit{rel=``next-memento''}) to the one that is returned. Note also the use of a HTTP Link header to point back to the Original Resource (\textit{rel=``original''}).


\section{Memento Resource Versioning}
\label{sec:mementoversion}

\begin{figure}[h!]
\begin{center}
\psfig{file=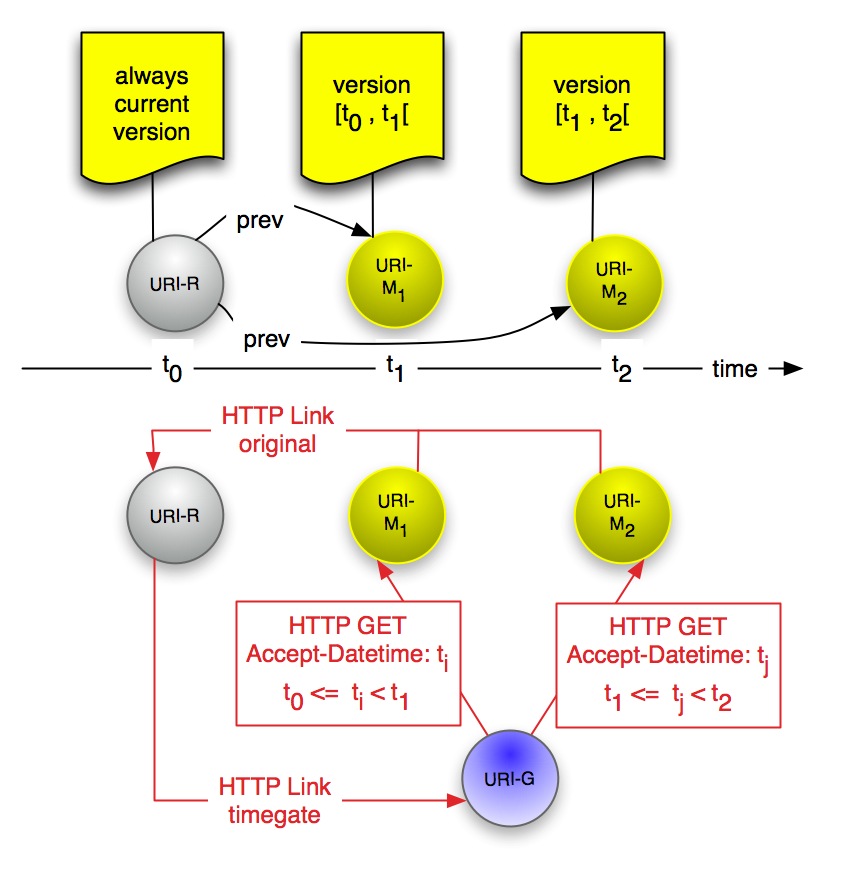, width=2.7in}
\caption{Memento Resource Versioning.} 
\label{fig:mementoversion}
\end{center}
\end{figure}

The Memento framework suggests a versioning mechanism that is fully based on HTTP (see Figure \ref{fig:mementoversion}). These are its core characteristics:

\begin{itemize}

\item \textit{Identification}: HTTP URIs are used to identify versions of resources. 

\item \textit{Versioning Strategy}: The top of Figure \ref{fig:mementoversion}
shows URI-R as the resource from which at any point in time the
current representation is served, and URI-M$_i$ as resources that
provide access to representations that were previously available from
URI-R. In terms of \cite{tbl:generic}
and its associated ontology\footnote{\url{http://www.w3.org/2006/gen/ont}},
URI-R is a \textit{time-generic} resource, whereas all URI-M$_i$ are
\textit{time-specific} resources. This strategy is different than the one
shown in Figure \ref{fig:classicversion}, yet aligned with the stable
URI principle of Cool URIs \cite{berners-lee:cool, coolURIssemanticweb}: instead of minting a
new URI for every \textbf{new} version, keep the URI stable
and mint new URIs for \textbf{old} versions. This approach
has become rather widespread; for example, \url{http://cnn.com} and
\url{http://en.wikipedia.org/wiki/DJ_Shadow} are such URI-R, whereas
\url{http://web.archive.org/web/20010911203610/http://www.cnn.com} and
\url{http://en.wikipedia.org/w/index.php?title=DJ_Shadow&oldid=337446696}
are examples of respective URI-M$_i$.

\item \textit{Version Relationships}: A \textit{timegate} HTTP Link header provided in response to GET/HEAD requests issued against the stable URI-R points at a TimeGate. And, an \textit{original} HTTP Link header provided in response to GET/HEAD requests issued at Mementos URI-M$_i$ points at URI-R of the Original Resource. As described, a TimeGate supports datetime conneg based on the content of the \textit{Accept-Datetime} header. It is a \textit{time-travel} resource that acts as a gateway between the time-generic URI-R and its associated time-specific Mementos URI-M$_i$. The result of the datetime conneg is a Memento that meets the expressed datetime preference. Also, the \textit{prev-memento} and \textit{next-memento} relationships may be used in the HTTP Link header to point at Mementos that are adjacent in time to the returned one.

\item \textit{Version Timestamping}: Versioning of URI-R is not required as it always is the current version. Mementos URI-M$_i$ are timestamped by means of the \textit{Content-Datetime} response header. 

\end{itemize}

The technology substrate used by the Memento versioning approach is fully centered on HTTP: URI, HTTP, HTTP Link with to-be-registered link relationships, HTTP datetime conneg. The challenges formulated in the Introduction can be addressed as follows (bottom of Figure \ref{fig:mementoversion}):

\begin{enumerate}

\item \textit{Given the current version of a resource, how can ``follow your nose'' style navigation to prior versions of the resource be achieved?}
The current version is the stable URI-R. A client application can follow its nose to a TimeGate for URI-R by using the URI that is expressed in the \textit{timegate} HTTP Link header returned by URI-R. The TimeGate supports datetime conneg allowing the client to obtain various versions (Mementos) by varying the content of its \textit{Accept-Datetime} request header. In addition, in cases where the \textit{prev-memento} and \textit{next-memento} relationships are available in the Alternates header provided in TimeGate responses, a client can engage in version-to-version navigation with a certainty that the version-relationships are time-based. 

\item \textit{Given any version of a resource and a particular timestamp, how can ``follow your nose'' style navigation towards another version that matches the timestamp be achieved?}
A version URI-M$_i$ provides an \textit{original} HTTP Link header pointing at the stable URI-R. From thereon, this scenario is the same as described in the previous point; the timestamp is used as the content of the \textit{Accept-Datetime} request header. The \textit{Content-Datetime} provides the earliest datetime at which the returned version became available; that version was still the then-current one at the datetime that was expressed in the \textit{Accept-Datetime} header.

\end{enumerate}

\section{Memento Resource Versioning and Linked Data}
\label{sec:linkeddata}

\begin{figure} 
\begin{center}
\psfig{file=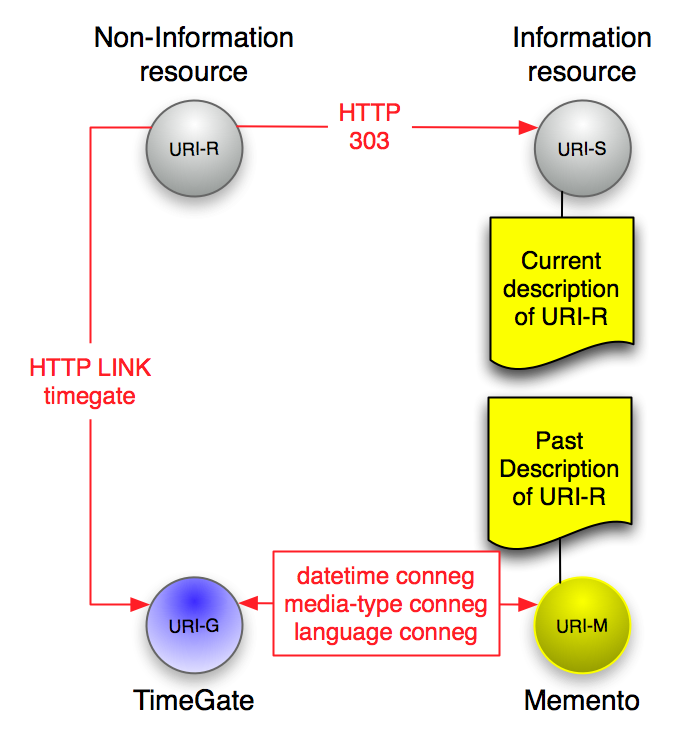, width=2.5in}
\caption{Memento Resource Versioning and Linked Data.} 
\label{fig:linkeddata}
\end{center}
\end{figure}

Figure \ref{fig:linkeddata} shows how Memento integrates in the Linked
Data environment. In this case, URI-R is a cool URI for a non-information
resource \cite{coolURIssemanticweb}, and the current description of URI-R
is available at URI-S. A TimeGate URI-G is introduced for URI-R, and to
support its discovery a \textit{timegate} HTTP Link header is provided
in responses to GET/HEAD requests to URI-R. When a Linked Data client
is in need of prior descriptions of URI-R, it follows its nose to URI-G,
where it can use datetime conneg to arrive at a description of URI-R as
it existed at some time in the past. Note that the conneg with URI-G
can include dimensions other than datetime. The media-type dimension
that is commonly used in Linked Data to allow a choice of descriptions
expressed in RDF serializations or HTML can also be supported. Similarly,
negotiation on language can be supported.

\section{The DBpedia Demonstrator}
\label{sec:dbpedia}

\subsection{Demonstrator Set-Up}
\label{sec:dbpedia-setup}

\begin{figure*}[ht]
\begin{center}
\includegraphics[scale=0.27]{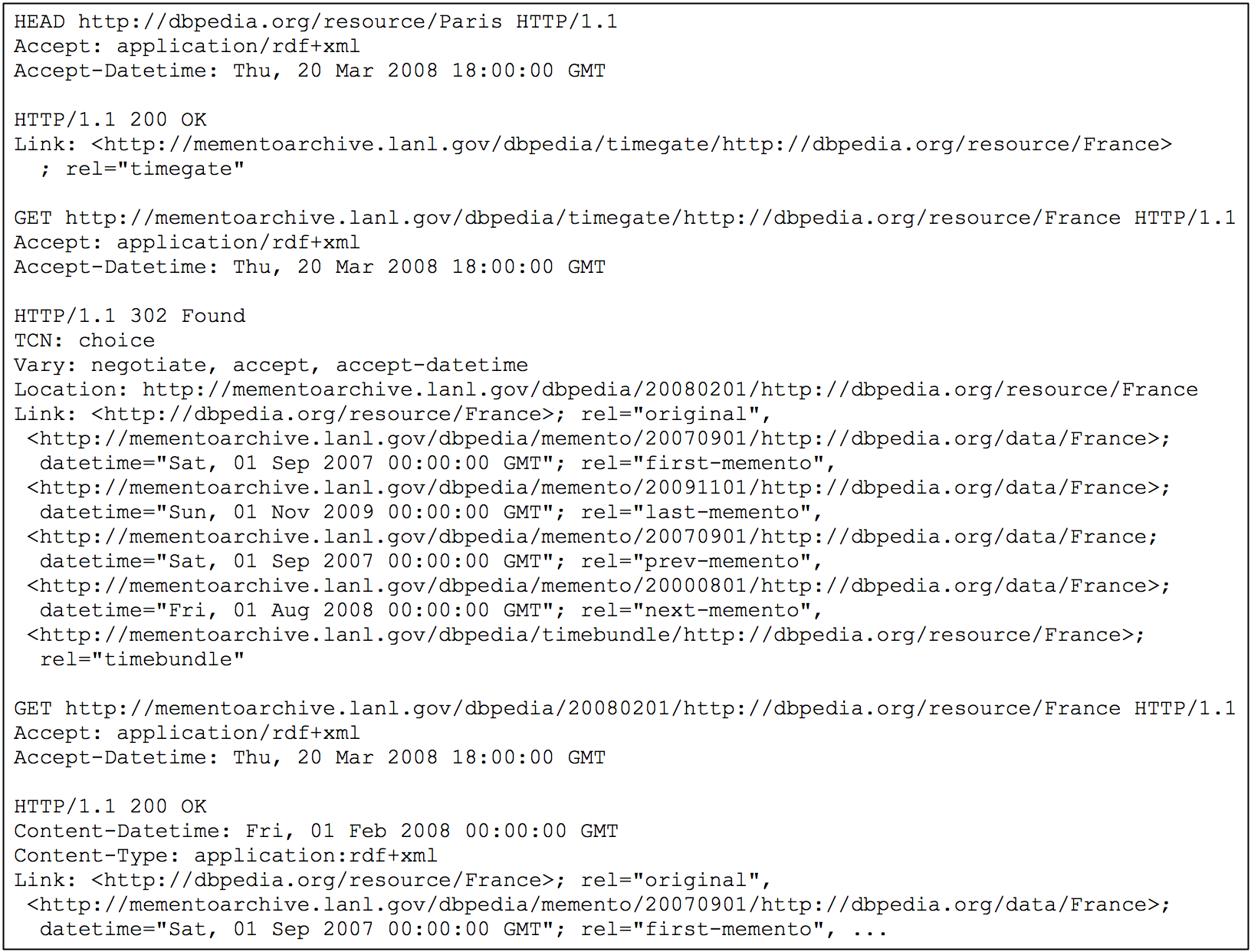}
\caption{Memento DBpedia HTTP Request/Response Cycle.}
\label{fig:linkeddataHTTP}
\end{center}
\end{figure*}



We have implemented the architecture depicted in Figure \ref{fig:linkeddata} in the DBpedia context.
We first downloaded the five prior English-language versions of DBpedia (2.0 through
3.3)\footnote{\url{http://wiki.dbpedia.org/Downloads34}} in NT
format. Using a python script, the approximately 600 Million triples were
loaded into a MySQL table (Table \ref{tab:dbase}). Loading
took approximately 15 hours, and resulted in a MyISAM table of 81 GB.

\begin{table} 
\begin{scriptsize}
\centering
\caption{DBpedia Demonstrator Database Table.}
\label{tab:dbase}
\begin{tabular}{|c|c|c|c|c|} 
\hline
id & subject & start & end & triples \\
\hline
integer, & varchar(256) & datetime & datetime & blob \\
auto\_increment, & & & & \\
not null, & & & & \\
PK & & & & \\
\hline
\end{tabular}
\end{scriptsize}
\end{table}


For each DBpedia subject URI-R, we exposed a TimeGate to support content negotiation in the datetime and media-type dimensions. For example, our TimeGate for DBpedia's France resource \url{http://dbpedia.org/resource/France} is \url{http://mementoarchive.lanl.gov/dbpedia/timegate/http://dbpedia.org/resource/France}. The datetime functionality was implemented by retrieving all distinct start/end combinations for the requested subject URI-R. This approach readily supports providing the \textit{first-memento}, \textit{last-memento}, \textit{next-memento} and \textit{prev-memento} relationships in the HTTP Link header provided in responses. The returned Memento is the one with the start/end interval that covers the datetime requested via conneg. For conneg requests with a datetime value that is in the range of the current DBpedia version, the TimeGate issues an HTTP 302 redirect to the Original Resource URI-R at \url{dbpedia.org}. Mementos are available both in HTML and RDF/XML. For example, the Memento for DBpedia 3.3's France resource in HTML is \url{http://mementoarchive.lanl.gov/dbpedia/memento/20090701/http://dbpedia.org/page/France}.
 
Colleagues at DBpedia kindly implemented the \textit{timegate} HTTP
Link header pointing at our TimeGates.  This required approximately
one hour and consisted of adding a stored procedure in the OpenLink
Virtuoso engine
to add the appropriate HTTP Link header.

Figure \ref{fig:linkeddataHTTP} shows a Memento HTTP flow
whereby a client requests the description of France that was available
from DBpedia on March 20, 2008. Only headers crucial to convey an
understanding of the conneg are shown. 

To illustrate full Memento compliance, we also implemented
TimeBundle/TimeGate support for our DBpedia version archive. This
functionality was not used to achieve the time-series analysis
described below. The Appendix shows our TimeMap for the DBpedia
resource \url{http://dbpedia.org/resource/France}; the content should
be self-explanatory.

Although DBpedia currently operates under a regime of recurrent discreet updates, both the proposed Memento approach and our specific database design support a possible future regime in which DBpedia is updated on an ongoing basis. In this case, an archiving mechanism would need to be added to ensure that versions of distinct DBpedia descriptions are pushed/pulled into the version archive as they change.


\subsection{Time-Series Analysis using Memento Resource Versioning}
\label{sec:timeseries}

To illustrate the power of the proposed approach, we implemented a
simple time-series analysis using both past and current DBpedia data.
We set out to trace the evolution over time of the Gross Domestic Product Per Capita for various countries, leveraging the \url{http://dbpedia.org/property/gdpPppPerCapita} property. 

The straightforward data-time-traveling algorithm used to construct the time-series data-set is described by the below pseudo code. It must be noted that the actual script must rely on some ad-hoc heuristics to deal with diverging data formats used for GDP values. 

\begin{scriptsize}
\begin{verbatim}
resources := [list of country description TimeGate URIs]
times := [list of date times, one per version including current]
prop := "http://dbpedia.org/property/gdpPppPerCapita"
values := {}

foreach r in resources:
   values[r] := []
   foreach t in times:
       data := fetch(URI-TG/r, Accept-Datetime: t, Accept:
"application/rdf+xml")
       graph := parse(data)
       value := graph.sparql(SELECT val WHERE { r prop ?val . })
       value := normalize(value)
       values[r].push(value)

\end{verbatim}
\end{scriptsize}

\medskip

\begin{figure*}
\begin{center}
\includegraphics[scale=0.59]{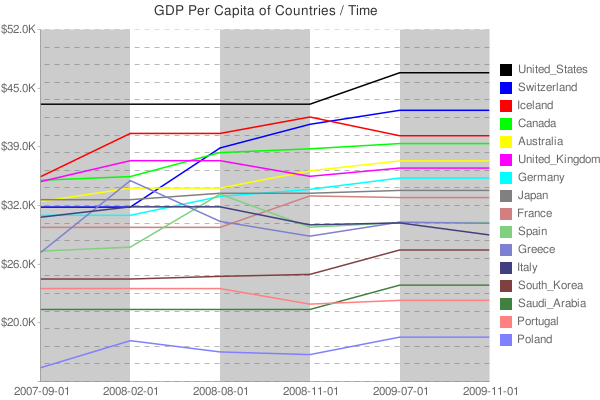}
\caption{A time-series analysis conducted across DBpedia versions using Memento's HTTP-based versioning approach.} 
\label{fig:timeseries}
\end{center}
\end{figure*}

The collected data were then turned into a chart
(Figure \ref{fig:timeseries}) using the Google Chart
API\footnote{\url{http://code.google.com/apis/charttools/index.html}}.

\section{Related Work}
\label{sec:related}

Little research has explored a protocol-based solution to augment the Web
with time travel capabilities. TTApache \cite{dyreson:managing} introduced
an ad-hoc RPC-style mechanism to access archived representations
given the URI of their original, e.g. ``page.html?02-Nov-2009''. This
approach reveals the local scope of the problem addressed by TTApache,
as opposed to the global perspective taken by the Memento datetime
conneg framework. Indeed, the query components are issued against a
specific server, and are not maintained when a client moves to another
server as is the case with the Accept-Datetime header of datetime
conneg. TTApache also allowed addressing archived representations
using version numbers in query components rather than datetimes. This
capability is similar to the deprecated ``Content-Version'' header
field from RFC 2068 \cite{rfc2068} and other, similar expired proposals
(e.g., \cite{meta-level-draft}). Such versioning features have not found
wide-spread adoption, presumably because their address space is tied to
a specific resource or server, and not universal like datetime. TTApache
also provided support for reserved terms as query components such as
``page.html?now''. This capability is similar to link relationship
types such as  ``latest-version'', ``predecessor-version'',
``successor-version'', and ``working-copy-of'' proposed in
\cite{rfc:link-relations}
to allow simple version navigation between Web resources. The
focus of this proposal that emerged from the AtomPub
\cite{rfc:5023} context, however, is clearly
on editorial version control (cf. WebDav, Java Content Repository). Also,
it provides no means to navigate versions based on datetime information.

There is a relationship between the described work and efforts that
research the problem of provenance of Linked Data, specifically those
provenance aspects concerned with the time intervals in which specific
data is valid. For example, \cite{hartig2009provenance}, is concerned
with provenance graphs that allow expressing such validity information,
whereas \cite{haslhofer2009dsnotify} focuses on applications to support
preserving link integrity over time. Our proposal introduces a native HTTP
approach that allows leveraging the results of these efforts at Web scale.

\section{Conclusions}
\label{sec:conclusion}

URIs like \url{http://weather.example.com/oaxaca} used in \cite{archWWW}
have gained significant functionality in the Linked Data context as they
start providing access not only to HTML intended for human consumption,
but also to data expressed in some RDF serialization intended for machine
processing. When publishing data in accordance to Memento's HTTP-based
versioning mechanism proposed in this paper, their value further increases
as they become entry points to both current and past versions of data. The
time-series analysis described in Section \ref{sec:timeseries}
is an admittedly simple demonstration of a subtle and powerful change in
the utility of Linked Data URIs.


The URI \url{http://weather.example.com/oaxaca} can now be leveraged
to obtain an overview of Oaxaca's weather in the past months, merely by
issuing HTTP GET requests with varying datetime preferences. Similarly,
time-traveling a Dow Jones data URI can result in an overview of the
stock market's evolution at any desired granularity. Tracing the evolving
state of traffic congestions, implemented in Zoetrope \cite{1449756}
by high-frequency crawls and scraping of a traffic web site could be
achieved by dereferencing a single data URI with varying timestamps instead. 

While this paper has focused on Linked Data, it should be clear that the proposed versioning mechanism can be applied to Web resources in general. It could, for example, be leveraged to facilitate navigating across issues of Web-based newspapers and magazines, and it can play an important role in better integrating the data-intensive eScience and eHumanities efforts into the Web. Hence, the addition of a time dimension to the Web is not something
only digital archaeologists should care about. It is an enabler for a
global HTTP-based versioning mechanism that can support a new range of
temporal applications for both the document and the data Web. This paper
has merely scratched the surface of a new world of possibilities.

\section{Acknowledgments}
The Memento research is partly funded by the Library of Congress. Many thanks to Chris Bizer, Kinsley Idehen, and Mitko Iliev for implementing the \textit{timegate} HTTP Link header in DBpedia.

\bibliographystyle{abbrv}
\bibliography{mln}  

\appendix
\label{sec:appa}
\section{An RDF/XML TimeMap} 

The following is an RDF/XML TimeMap for \url{http://dbpedia.org/resource/France}.

\begin{scriptsize}
\begin{verbatim}
<?xml version="1.0" encoding="utf-8"?>
<rdf:RDF
  xmlns:foaf='http://xmlns.com/foaf/0.1/'
  xmlns:dcterms='http://purl.org/dc/terms/'
  xmlns:mem='http://www.mementoweb.org/terms/tb/'
  xmlns:dc='http://purl.org/dc/elements/1.1/'
  xmlns:rdf='http://www.w3.org/1999/02/22-rdf-syntax-ns#'
  xmlns:ore='http://www.openarchives.org/ore/terms/'
  xmlns:rdfs1='http://www.w3.org/2001/01/rdf-schema#'>
<ore:ResourceMap rdf:about="http://mementoarchive.lanl.gov/dbpedia/timemap/rdf/http://dbpedia.org/resource/France">
<rdf:type rdf:resource="http://www.mementoweb.org/terms/tb/TimeMap"/>
<dcterms:modified>2010-02-17T05:26:27Z</dcterms:modified>
<dcterms:created>2010-02-17T05:26:27Z</dcterms:created>
<dc:format>application/rdf+xml</dc:format>
<dcterms:creator>
<rdf:Description rdf:about="http://foresite-toolkit.googlecode.com/#pythonAgent">
<foaf:mbox>foresite@googlegroups.com</foaf:mbox>
<foaf:name>Foresite Toolkit (Python)</foaf:name>
</rdf:Description>
</dcterms:creator>
<ore:describes>
<ore:Aggregation rdf:about="http://mementoarchive.lanl.gov/dbpedia/timebundle/http://dbpedia.org/resource/France">
<ore:aggregates rdf:resource="http://mementoarchive.lanl.gov/dbpedia/memento/20070901/http://dbpedia.org/data/France"/>
<ore:aggregates rdf:resource="http://mementoarchive.lanl.gov/dbpedia/memento/20080201/http://dbpedia.org/data/France"/>
<ore:aggregates rdf:resource="http://mementoarchive.lanl.gov/dbpedia/memento/20080801/http://dbpedia.org/data/France"/>
<ore:aggregates rdf:resource="http://mementoarchive.lanl.gov/dbpedia/memento/20081101/http://dbpedia.org/data/France"/>
<ore:aggregates rdf:resource="http://mementoarchive.lanl.gov/dbpedia/memento/20090701/http://dbpedia.org/data/France"/>
<ore:aggregates rdf:resource="http://mementoarchive.lanl.gov/dbpedia/timegate/http://dbpedia.org/resource/France"/>
<ore:aggregates rdf:resource="http://dbpedia.org/resource/France"/>
<dc:title>Memento Time Bundle for http://dbpedia.org/resource/France</dc:title>
<rdf:type rdf:resource="http://www.mementoweb.org/terms/tb/TimeBundle"/>
</ore:Aggregation>
</ore:describes>
</ore:ResourceMap>
























































































  <rdf:Description rdf:about="http://www.openarchives.org/ore/terms/Aggregation">
    <rdfs1:label>Aggregation</rdfs1:label>
    <rdfs1:isDefinedBy rdf:resource="http://www.openarchives.org/ore/terms/"/>
  </rdf:Description>
  <rdf:Description rdf:about="http://www.openarchives.org/ore/terms/ResourceMap">
    <rdfs1:label>ResourceMap</rdfs1:label>
    <rdfs1:isDefinedBy rdf:resource="http://www.openarchives.org/ore/terms/"/>
  </rdf:Description>
  <mem:TimeGate rdf:about="http://mementoarchive.lanl.gov/dbpedia/timegate/http://dbpedia.org/resource/France">
    <mem:timeGateFor rdf:resource="http://dbpedia.org/resource/France"/>
    <mem:covers rdf:nodeID="gtKXRtug21"/>
  </mem:TimeGate>
 <mem:Period rdf:nodeID="gtKXRtug21">
    <mem:end rdf:datatype="http://www.w3.org/2001/XMLSchema#dateTime">2009-11-01T00:00:00+00:00</mem:end>
    <mem:start rdf:datatype="http://www.w3.org/2001/XMLSchema#dateTime">2007-09-01T00:00:00+00:00</mem:start>
  </mem:Period>
  <mem:Memento rdf:about="http://mementoarchive.lanl.gov/dbpedia/memento/20070901/http://dbpedia.org/data/France">
    <mem:validOver>
      <mem:Period>
        <mem:start rdf:datatype="http://www.w3.org/2001/XMLSchema#dateTime">2007-09-01T00:00:00+00:00</mem:start>
        <mem:end rdf:datatype="http://www.w3.org/2001/XMLSchema#dateTime">2008-02-01T00:00:00+00:00</mem:end>
      </mem:Period>
    </mem:validOver>
    <mem:mementoFor rdf:resource="http://dbpedia.org/resource/France"/>
  </mem:Memento>
  <mem:Memento rdf:about="http://mementoarchive.lanl.gov/dbpedia/memento/20080201/http://dbpedia.org/data/France">
    <mem:validOver>
      <mem:Period>
        <mem:start rdf:datatype="http://www.w3.org/2001/XMLSchema#dateTime">2008-02-01T00:00:00+00:00</mem:start>
        <mem:end rdf:datatype="http://www.w3.org/2001/XMLSchema#dateTime">2008-08-01T00:00:00+00:00</mem:end>
      </mem:Period>
    </mem:validOver>
    <mem:mementoFor rdf:resource="http://dbpedia.org/resource/France"/>
  </mem:Memento>
  <mem:Memento rdf:about="http://mementoarchive.lanl.gov/dbpedia/memento/20080801/http://dbpedia.org/data/France">
    <mem:validOver rdf:nodeID="gtKXRtug42"/>
    <mem:mementoFor rdf:resource="http://dbpedia.org/resource/France"/>
  </mem:Memento>
  <mem:Period rdf:nodeID="gtKXRtug42">
    <mem:end rdf:datatype="http://www.w3.org/2001/XMLSchema#dateTime">2008-08-01T00:00:00+00:00</mem:end>
    <mem:start rdf:datatype="http://www.w3.org/2001/XMLSchema#dateTime">2009-11-01T00:00:00+00:00</mem:start>
  </mem:Period>
  <mem:Memento rdf:about="http://mementoarchive.lanl.gov/dbpedia/memento/20081101/http://dbpedia.org/data/France">
    <mem:validOver rdf:nodeID="gtKXRtug49"/>
    <mem:mementoFor>
      <mem:OriginalResource rdf:about="http://dbpedia.org/resource/France"/>
    </mem:mementoFor>
  </mem:Memento>
  <mem:Period rdf:nodeID="gtKXRtug49">
    <mem:end rdf:datatype="http://www.w3.org/2001/XMLSchema#dateTime">2008-11-01T00:00:00+00:00</mem:end>
    <mem:start rdf:datatype="http://www.w3.org/2001/XMLSchema#dateTime">2009-07-01T00:00:00+00:00</mem:start>
  </mem:Period>
  <mem:Memento rdf:about="http://mementoarchive.lanl.gov/dbpedia/memento/20090701/http://dbpedia.org/data/France">
    <mem:validOver rdf:nodeID="gtKXRtug56"/>
    <mem:mementoFor rdf:resource="http://dbpedia.org/resource/France"/>
  </mem:Memento>
  <mem:Period rdf:nodeID="gtKXRtug56">
    <mem:end rdf:datatype="http://www.w3.org/2001/XMLSchema#dateTime">2009-07-01T00:00:00+00:00</mem:end>
    <mem:start rdf:datatype="http://www.w3.org/2001/XMLSchema#dateTime">2009-11-01T00:00:00+00:00</mem:start>
  </mem:Period>
</rdf:RDF>
\end{verbatim}
\end{scriptsize}


\end{document}